\documentclass[aps,prd,10pt,nofootinbib,onecolumn,eqsecnum,showpacs,showkeys,superscriptaddress,preprintnumbers,altaffilletter]{revtex4}
\usepackage{graphicx}
\usepackage{amsmath}
\usepackage{multirow}
\usepackage{hyperref} 
\usepackage{soul}
\usepackage{graphicx}
\usepackage{dcolumn}
\usepackage{amssymb}
\usepackage{amsfonts}
\usepackage{amsbsy}
\usepackage{color}
\usepackage{rotating}
\usepackage[english]{babel}
\usepackage{multirow}
\usepackage{float}
\usepackage{cleveref}
\usepackage{diagbox} 
\usepackage{standalone} 
\usepackage{import}
\usepackage{tikz} 
\tikzset{
  pics/carc/.style args={#1:#2:#3}{
    code={
      \draw[pic actions] (#1:#3) arc(#1:#2:#3);
    }
  }  
}

\usepackage{booktabs}

\newcolumntype{x}[1]{>{\centering\arraybackslash\hspace{0pt}}p{#1}}

\newcommand{\be}{\begin{equation}}
\newcommand{\ee}{\end{equation}}
\newcommand{\bea}{\begin{eqnarray}}
\newcommand{\eea}{\end{eqnarray}}
\newcommand{\beal}{\begin{aligned}}
\newcommand{\eeal}{\end{aligned}}
\newcommand{\bi}{\begin{itemize}}
\newcommand{\ei}{\end{itemize}}

\hypersetup{
    colorlinks=true,
    linkcolor=blue,
    filecolor=magenta,
    urlcolor=cyan,
    citecolor=cyan
}

\begin{document}

\title{Generalized Nonextensive Entropy Holographic Dark Energy Models Verified by Cosmological Data} 
\date{today}

\author{Ilim \c{C}imdiker}
\email{ilim.cimdiker@phd.usz.edu.pl}
\affiliation{Institute of Physics, University of Szczecin, Wielkopolska 15, 70-451 Szczecin, Poland}
\author{Mariusz P. D\c{a}browski}
\email{mariusz.dabrowski@usz.edu.pl}
\affiliation{Institute of Physics, University of Szczecin, Wielkopolska 15, 70-451 Szczecin, Poland}
\affiliation{National Centre for Nuclear Research, Andrzeja So{\l}tana 7, 05-400 Otwock, Poland}
\affiliation{Copernicus Center for Interdisciplinary Studies, Szczepa\'nska 1/5, 31-011 Krak\'ow, Poland}
\author{V. Salzano}
\email{vincenzo.salzano@usz.edu.pl}
\affiliation{Institute of Physics, University of Szczecin, Wielkopolska 15, 70-451 Szczecin, Poland}

\date{\today}

\begin{abstract}

We present a general formalism for studying generalized Holographic Dark Energy (HDE) models in which we use a dimensionless form of the area-entropy of cosmological horizons. The future event horizon is applied though the formalism can also be applied to any other type of the horizon, too. Then, we use our formalism for nonextensive horizon entropies of standard HDE (i.e. Bekenstein-Hawking), and generalized such as Barrow/Tsallis-Cirto, R\'enyi, Sharma-Mittal, and Kaniadakis as dark energy models of the universe and test them by cosmological data. 

We find the bounds on the specific entropy model parameters and also apply statistical comparison tool such as the Bayesian evidence criterion in order to favour or disfavour the models against standard $\Lambda$CDM. 

The main data test results are that all the HDE models under study are statistically disfavoured with respect to $\Lambda$CDM, though at some different levels. The standard HDE seem to be on the same footing as R\'enyi, Sharma-Mittal, and Kaniadakis HDE models since the latter include only small deviations from HDE model resulting from the series expansion of their extra nonextensivity parameters. However, Barrow and Tsallis-Cirto models, though still disfavoured against $\Lambda$CDM, seem to point out observationally to fulfil an important physical property of extensivity (though still remaining nonadditive).
Finally, the Tsallis-Cirto model parameter is pointing towards the $\Lambda$CDM limit which is singular also at the expense of having much larger value of the holographic dark energy dimensionless parameter $k$ value higher than other models. 

\end{abstract}

\maketitle

\section{Introduction}

In the standard Boltzmann-Gibbs thermodynamics and statistical mechanics the entropy is assumed an extensive quantity, coming as a result of the negligence of the long-range interactions between thermodynamic subsystems. Here, the size of a system exceeds the range of the interaction between its components leading to the total entropy of a composite system to be equal to the sum of the entropies of individual subsystems (resulting in additivity) and the entropy grows with the size of the system or its configuration space (resulting in extensivity). 

However, extensivity is not a property of gravitational systems where the long-range forces are important, and besides, gravity is strongly non-linear when its extreme regimes are taken into account. Strong gravity is a characteristic of the compact objects in the Universe including the most extreme of them - the black holes, which are surrounded by the horizons which areas, according to Bekenstein and Hawking \cite{Bekenstein:1973ur,Hawking:1975vcx}, are interpreted as their entropy. Black holes with the Bekenstein entropy scales with the area and not with the volume (size), and so is consequently a nonextensive quantity \cite{Tsallis:2012js,Biro:2013cra,Czinner:2017bwc,Czinner:2015eyk,Czinner:2017tjq,Tsallis:2019giw}. 
 
The plethora of nonextensive entropies have been proposed in the literature \cite{Tsallis:1987eu,Renyi1959,SM,sharma1975new,Kaniadakis:2002zz,Kaniadakis:2005zk,Barrow:2020tzx} (for a recent review see \cite{Entropy2024}). Recently, a number of them have been applied to gravitational systems as candidates for dark energy \cite{Saridakis:2020zol,Dabrowski:2020atl,Saridakis:2020lrg,AlMamon:2020usb,Asghari:2021lzu,Adhikary:2021xym,Nojiri:2021czz,Drepanou:2021jiv,DiGennaro:2022ykp,DiGennaro:2022grw,Nojiri:2021jxf,Anagnostopoulos:2020ctz,Sheykhi:2018dpn,FeiziMangoudehi:2022rwj,Ghoshal:2021ief,Luciano2022a,Leon:2021wyx,Almada2022_Kaniad_test}, which is  
 dubbed the holographic dark energy - HDE (see \cite{Wang:2016och}). 

In this paper, we apply a number of nonextensive entropies as holographic screens of the universe and confront them with observational data. Our focus will be put on: Barrow $\Delta$-entropy \cite{Barrow:2020tzx}, Tsallis-Cirto $\delta$-entropy \cite{Tsallis:2012js}, R\'enyi entropy \cite{Renyi1959}, Sharma-Mittal entropy  \cite{sharma1975new,SM}, and Kaniadakis entropy \cite{Kaniadakis:2002zz,Kaniadakis:2005zk}. 

The following is the outline of the paper. In Sec. \ref{GenHorEntropy}, we generalize the prospective dark energy density onto an arbitrary holographic dark energy model. In Sec. \ref{GenHDE} we present cosmological evolution equations which can be applied to any nonextensive entropy model. In Sec. \ref{appl_nonext} we apply these general equations to the above mentioned entropies in order to finally test them in Sec. \ref{nonext_v_data} by the full set of geometrical data in cosmology. Finally, in Sec. \ref{Discussion}, we present the results of our studies.

\section{Generalizing the energy density onto an arbitrary nonextensive horizon entropy model}
\label{GenHorEntropy}

The dark energy density in the standard $\Lambda$CDM model is related to the cosmological constant $\Lambda$ (with unit $m^{-2}$) as follows \cite{Amendola_Tsujikawa_2010}
\begin{equation}
    \rho_{\Lambda}= \frac{\Lambda c^2}{8 \pi G} ,
    \label{Lambda}
\end{equation}
where $c$ is the speed of light, $G$ is the gravitational constant, and $\rho_{\Lambda}$ is the mass density in units $kg \cdot m^{-3}$. It suffers from the hierarchy problem which says that its interpretation as vacuum energy gives its order of magnitude of 120 orders of magnitude higher than an observational bound \cite{PeeblesRatra2003} and so a lot of other than $\Lambda$CDM models have been proposed. Since the effect of $\Lambda$ is detectable gravitationally, only on the large scale (as an effect of cosmic acceleration), and its physical roots are unknown - this is why it is called the dark energy. The latter issue made cosmologists looking for some physical models which can explain the effect. 

A fraction of these models are based on the contribution to the energy density from the cosmological horizons in analogy to the black hole horizons which come in relation to the horizon entropy and are known as holographic dark energy (HDE) \cite{Wang:2016och}. We generalize it by introducing an arbitrary function $S(L)$ of the cosmological horizon $L$ representing a dimensionless entropy, as below  
\begin{equation}
    \rho_{HDE}=\frac{3 c^2k^2L_0^2}{8 \pi G} S(L) L^{-4} ,
    \label{HDE}
\end{equation}
where $k$ is a dimensionless constant related to the holographic screen properties \cite{Wang:2016och},  and 
\be 
L_0^2 = 4 G \hbar /c^3 = A_0
\label{L_0}
\ee 
is the Planck area with $L_0$ being of the size of the Planck length $l_p = 2 \sqrt{\hbar G/c^3}$. In (\ref{HDE}) the quantities have been chosen in a way that $\rho_{HDE}$ is given in units of mass density $kg \cdot m^{-3}$ in order to agree with the units of (\ref{Lambda}).  In other words, both (\ref{Lambda}) and (\ref{HDE}) have unit of mass density as it is expected in cosmology. 

In our approach, we use a number of nonextensive entropies which go beyond the standard extensive and additive Boltzmann-Gibbs entropy. The simplest case of such a nonextensive entropy is the Bekenstein-Hawking entropy \cite{Bekenstein:1973ur,Hawking:1975vcx} which given by
\begin{equation}
S_{BH} = 4 \pi k_B \left( \frac{M}{m_p} \right)^2 = \frac{4 \pi k_B G M^2} {\hbar c}  ,
\label{BH}
\end{equation}
where $k_B$ is the Boltzmann constant, $m_p = \hbar c/G$ is the Planck mass, $\hbar$ is the reduced Planck constant, and $M$ is the black hole mass. Bekenstein-Hawking entropy (\ref{BH}) can be fitted to our scheme of a dimensionless entropy (\ref{HDE}) by choosing 
\begin{equation}
    S_{BH}(L)=\frac{S_{BH}}{k_B} \equiv  \frac{A}{A_0} = \left( \frac{L}{L_0} \right)^2 = \frac{c^3}{4 G \hbar} L^2  ,
    \label{SL}
\end{equation}
where we have defined the horizon area as $A \propto L^2$ and the Planck area $A_0 = L_0^2$, as given by (\ref{L_0}). From (\ref{SL}), it is clear that in the limit $S(L)\rightarrow S_{BH}$, one obtains the standard holographic dark energy density as in Ref. \cite{Wang:2016och}, i.e., 
\be
 \rho_{HDE}=\frac{3 c^2k^2}{8 \pi G} L^{-2}
    \label{stHDE}
\ee
where the $C$ parameter introduced in Ref. \cite{Wang:2016och} is replaced by $k$ which is expected to take values of the order of one.

The motivation for introducing the dimensionless entropy (\ref{SL}) and dimensionless constant $k$ comes from two sides. Firstly, similarly as the coupling constants in particle physics, dimensionless quantities are universal since they are not subject to scale transformations and allow the comparison of various forms of entropies under study in a consistent way. In our approach (cf. (\ref{SL})), it comes naturally as the ratio of the cosmological horizon area $A$ to the minimum area $A_0$ given by the Planck length in (\ref{L_0}). This exactly fits the common normalization of the entropy (area) in the forms of elementary bits of information which is applied in the quantum theory of information or in quantum gravity framework nowadays \cite{Egan_2010}. Secondly, the introduction of the dimensionless constant $k$ allows to avoid confusion in having it dimensionful which causes some ugliness while handling the comparison with data. We found such an approach more convenient and appropriate for various entropies studies despite the motivations for these entropies come from very different physical backgrounds as described for example in Ref. \cite{Entropy2024}. 

\section{A general holographic dark energy model}
\label{GenHDE}

Our starting point is the holographic dark energy density model as given by (\ref{HDE}), which after following the standard cosmological modelling procedure \cite{Amendola_Tsujikawa_2010}, gives a dimensionless energy density as follows: 
\begin{equation}
 \Omega_{HDE} (a) = \frac{\rho}{\rho_{crit}} = \frac{c^2 k^2 L_0^2}{H^2(a)} S\left(L(a)\right) L^{-4}(a) , 
 \label{OmHDE}
\end{equation}
where we have taken $\rho_{crit}=3 H^2/ 8\pi G$ with $a(t)$ the cosmological scale factor and $H(t) = \dot{a}(t)/a(t)$ the Hubble parameter - the dot means the derivative with respect to cosmic time $t$. 
In (\ref{OmHDE}), we have also taken into account the fact that the size of the cosmological horizon $L$ changes in time according to the evolution of the scale factor $a$, and so it is the function of the scale factor $L=L(a)$. 

The other dimensionless parameters for the matter (m) and radiation (r) read 
\begin{equation}
    \Omega_m(a)= \Omega_{m0} a^{-3} \frac{H_0^2}{H^2(a)}\;,\; \Omega_r(a)= \Omega_{r0} a^{-4} \frac{H_0^2}{H^2(a)} ,
    \label{Omegas}
\end{equation}
where label '0' refers to the current epoch of the evolution. By the assumption of the flat cosmological geometry which requires 
\begin{equation}
    \Omega_m + \Omega_r + \Omega_{HDE} =1 ,
\end{equation}
we can write the following relation 
\begin{equation}
    1-\Omega_{HDE}(a) = \frac{H_0^2}{H^2(a)} (\Omega_{m0} a^{-3}+\Omega_{r0}a^{-4}).
    \label{1Omega}
    \end{equation}
Defining 
\begin{equation}
    \Phi(a) \equiv \Omega_{m0} a^{-3}+\Omega_{r0}a^{-4} ,
    \label{Phi}
\end{equation}
and taking the derivative of (\ref{Phi}) with respect to the scale factor $a$, we can write 
\be
\frac{\Phi^{\prime}}{\Phi} = - \frac{1}{a} \frac{ 3\Omega_{m0} a + 4 \Omega_{r0} }{ \Omega_{m0} a +\Omega_{r0} } ,
\label{PhiPr}
\ee
where $(\ldots)^\prime \equiv d/da$.

The dimensionless energy densities (\ref{Omegas}) can also be expressed as the function of the cosmological redshift  $z$ as follows
\be 
    \Omega_m = \frac{H_0^2}{H^2(z)} \Omega_{m0} (1+z)^3  , \hspace{0.3cm} \Omega_r = \frac{H_0^2}{H^2(z)} \Omega_{r0} (1+z)^4 , 
    \label{OmMR}
\ee
 where the scale factor is expressed as $a(z) = a_0 (1+z)^{-1}$ under the assumption that the current value of it $a(z=0) \equiv a_0 = 1$.  

Combining (\ref{OmHDE}) and (\ref{1Omega}) yields;
\begin{equation}
\frac{S(L(a))}{(L(a)/L_0)^4}= {\cal A}^2 \Phi(a)  \times \frac{\Omega_{HDE}(a)}{1-\Omega_{HDE}(a)} \equiv \mathcal{F}(\Omega_{HDE}(a))
\label{omeq}
\end{equation}
where we have defined a dimensionless constant
\be
{\cal A}^2 \equiv \frac{H_0^2 L_0^2}{k^2 c^2} .
\ee

In order to calculate the evolution of dark energy, we need to take the derivative of (\ref{omeq}) with respect to the scale factor $a$ splitting it for the left-hand side (LHS) and the right-hand side (RHS), as follows: 
\begin{equation}
    LHS= \left(\frac{L_0}{L}\right)^4 S(L) \left(  \frac{ \frac{\partial S(L)}{\partial L}}{S(L)} L^{\prime}  - 4 \frac{L^{\prime}}{L} \right) 
    \label{OmLHS}
\end{equation}
\begin{equation}
    RHS= \frac{\Phi^{\prime}(a)}{\Phi(a)}  S(L) \left(\frac{L_0}{L}\right)^4  + {\cal A}^2 \Phi \frac{\Omega_{HDE}^{\prime}}{\left( 1 - \Omega_{HDE} \right)^2} , 
    \label{OmRHS}
\end{equation}
where  (\ref{omeq}) have been applied. 
Combining (\ref{OmHDE}, (\ref{OmLHS}), and (\ref{OmRHS}) yields
\begin{equation}
\frac{\Omega_{HDE}'}{\Omega_{HDE}(\Omega_{HDE}-1)} =    \frac{\Phi'(a)}{\Phi(a)}  + 4 \frac{L^{\prime}}{L} -  \frac{\frac{\partial S(L)}{\partial L}}{S(L)} L^{\prime}   ,
\label{OmPrime}
\end{equation}

Although in general there are a couple of options for the selection of the horizons, from now on we will take the future event horizon radius defined as \cite{Dabrowski:2020atl}
\begin{equation}
    L(a) = c a \int_a^\infty \frac{da}{H a^2} ,
\end{equation}
with its derivative with respect to the scale factor $a$ given by
\begin{equation}
    \frac{dL}{da}= L^{\prime} = \frac{L}{a}-\frac{c}{a H} 
    \label{lprime} ,
\end{equation}
which applied to (\ref{OmPrime}) gives
\begin{equation}
  a \frac{\Omega_{HDE}'}{\Omega_{HDE}(\Omega_{HDE}-1)}=   a \frac{\Phi'(a)}{\Phi(a)} +  \left(1-\frac{c}{L H}\right) \left( 4 - \frac{\frac{\partial S(L)}{\partial L}}{S(L)} L  \right) .
\end{equation}
Finally, by using (\ref{OmHDE}), we obtain
\begin{equation}
\Omega_{HDE}'=\frac{\Omega_{HDE}(\Omega_{HDE}-1)}{a}\left[a \frac{\Phi'}{\Phi}+\left(1-\frac{\sqrt{\Omega_{HDE}}}{k} \times \sqrt{\frac{(L(a)/L_0)^2}{S}}\right)\times \left(4-\frac{\frac{\partial S(L)}{\partial L}}{S(L)} L \right) \right]  ,
\label{OmPrimeFin}  
\end{equation}
which is the most general evolution equation that we can get since it is independent of a choice of an explicit form of the entropy function $S(L)$.

\section{Application of nonextensive entropy holographic dark energy models}
\label{appl_nonext}

In our previous papers, we have studied various nonextensive entropies impact on black hole horizons \cite{Cimdiker:2022ics,Cimdiker2023}. In this paper we extend the discussion onto cosmological horizons, following parallel studies of the topic by us \cite{Dabrowski:2020atl,PhysRevD.108.103533}, and many other authors \cite{Saridakis:2020zol,Dabrowski:2020atl,Saridakis:2020lrg,AlMamon:2020usb,Asghari:2021lzu,Adhikary:2021xym,Nojiri:2021czz,Drepanou:2021jiv,DiGennaro:2022ykp,DiGennaro:2022grw,Nojiri:2021jxf,Anagnostopoulos:2020ctz,Sheykhi:2018dpn,FeiziMangoudehi:2022rwj,Ghoshal:2021ief,Luciano2022a,Leon:2021wyx,Almada2022_Kaniad_test}. 

\subsection{Bekenstein-Hawking nonextensive horizon entropy}

The most common form of the horizon entropy which serves as holographic dark energy is the Bekenstein-Hawking entropy (\ref{SL}) which after neglecting radiation $\Omega_{r0} =0$ (then Eq. (\ref{PhiPr}) gives the value of $\Phi^{\prime}/\Phi = -3/a$) produces the evolution equation in the form 
\begin{equation}
a \frac{\Omega_{HDE}'}{\Omega_{HDE}(1- \Omega_{HDE})} = 1 + \frac{2}{k} \sqrt{\Omega_{HDE}} 
\end{equation}
and it agrees with the formula (41) of Ref. \cite{Wang:2016och}, when our constant $k$ is replaced by the constant $C$. In fact, the Bekenstein-Hawking type nonextensive entropy as applied to cosmology is nothing else than the original form of HDE of Ref. \cite{Wang:2016och}.

\subsection{Barrow/Tsallis-Cirto nonextensive horizon entropy}

Barrow entropy is defined as \cite{Barrow:2020tzx}
\begin{equation}
    S_B(L)=\left(\frac{L}{L_0}\right)^{2+\Delta} ,
    \label{SBar}
\end{equation}
where $\Delta$ is a fractality parameter from the range $0 \leq \Delta \leq 1$. Using (\ref{SBar}), the function $\mathcal{F}$ in (\ref{omeq}) reads 
\be
\mathcal{F}_B(\Omega_{HDE}(a)) =  \left(\frac{L}{L_0}\right)^{\Delta - 2} , 
\label{F_Barrow}
\ee
which applied to the evolution of dark energy equation (\ref{OmPrimeFin}) gives 
 \be
\Omega_{HDE}' = \frac{\Omega_{HDE}(\Omega_{HDE}-1)}{a} \left\{  a \frac{\Phi'(a)}{\Phi(a)}  + (2 - \Delta)  \left[1- \frac{\sqrt{\Omega_{HDE}} }{k} \mathcal{F}_B^{\frac{\Delta}{2(2-\Delta)}}(\Omega_{HDE}(a))    \right] \right\} ,
\label{HDEBar1}
\ee
and further using (\ref{PhiPr}) 
gives
 \bea
&& a \frac{\Omega_{HDE}'}{\Omega_{HDE}(1 - \Omega_{HDE})} =  \frac{(\Delta +1) \Omega_{m0} a + (\Delta +2) \Omega_{r0} }{\Omega_{m0} a + \Omega_{r0} }  + \nonumber \\
&+& (2 - \Delta)  k^{\frac{2}{\Delta -2}} \left(\frac{c}{H_0L_0} \right)^{\frac{\Delta}{\Delta -2}}  \left(\Omega_{HDE}\right)^{\frac{1}{2-\Delta}} \left(1 - \Omega_{HDE} \right)^{\frac{\Delta}{2(\Delta-2)}} \left(\Omega_{m0} a + \Omega_{r0} \right)^{\frac{\Delta}{2(2-\Delta)}} a^{\frac{2\Delta}{\Delta - 2}} .
\label{HDEBarfin}
\eea
The formula (\ref{HDEBarfin}) agrees with the formula (3.11) of Ref. \cite{PhysRevD.108.103533}, provided we define the relation of the dimensionful holographic constant $C$ from \cite{PhysRevD.108.103533} with the dimensionless parameter $k$ of ours as 
\be
C=ck L_0^{-\Delta/2}  .
\label{Ck}
\ee
Note that for explicit values of $c$ and $L_o$, (\ref{Ck}) is equal to $C = 3\cdot 10^{2(4+9\Delta)} 2^{-2\Delta} k $ in units of $ m^{1-\Delta/2} s^{-1}$, and in the extreme case of $\Delta =1$, it gives $C \sim 0.75 \cdot 10^{26} k$ in units $m^{1/2} s^{-1}$.  The Barrow entropy evolution equation (\ref{HDEBarfin}) is limited to the fractality parameter range $0 \leq \Delta \leq 1$. This actually prevents the $\Lambda$CDM model limit $\Delta \to 2$. More general entropy definition which can be extended beyond the Barrow range of $\Delta$ is the Tsallis-Cirto entropy \cite{Tsallis:2012js}, and it can be obtained by replacing the fractality parameter $\Delta$ by a nonextensivity parameter $\delta$ as 
\be
\Delta = 2 (\delta -1) , 
\label{Deldel}
\ee
which gives the dark energy evolution equation (\ref{HDEBarfin}) in the form 
 \bea
&& a \frac{\Omega_{HDE}'}{\Omega_{HDE}(1 - \Omega_{HDE})} =  \frac{(2 \delta -1) \Omega_{m0} a + 2\delta \Omega_{r0} }{\Omega_{m0} a + \Omega_{r0} }  + \nonumber \\
&+& 2(2 - \delta)  k^{\frac{1}{\delta -2}} \left(\frac{c}{H_0L_0} \right)^{\frac{\delta-1}{\delta -2}}  \left(\Omega_{HDE}\right)^{\frac{1}{2(2-\delta)}} \left(1 - \Omega_{HDE} \right)^{\frac{\delta-1}{2(\delta-2)}} \left(\Omega_{m0} a + \Omega_{r0} \right)^{\frac{\delta-1}{2(2-\delta)}} a^{\frac{2(\delta-1}{\delta - 2}} .
\label{HDETsallisC}
\eea
In principle, $\delta$ can take any real value, though due to the requirement of concavity of entropy (having a minimum), it usually is taken to be positive i.e. $\delta > 0$. It is also worth noticing that the limit $\delta =2$ (which is equal to the limit $\Delta =2$ - cf. (\ref{Deldel})), is excluded from the range of calculations in (\ref{HDETsallisC}). This limit would in principle be reproducing the standard $\Lambda$CDM model. 

As it is clear from Sections \ref{GenHorEntropy} and \ref{GenHDE} we apply holographic principle which means that we want to replace the cosmological constant {\it fully} by HDE. This is different from the gravity-thermodynamics conjecture (see e.g. Ref. \cite{Lymperis2018}). where the cosmological constant is already present and the horizon entropy contribution is only a small correction to the standard $\Lambda$CDM model. This means that we can have HDE mimicking the cosmological constant in some limit. This limit is easily identified as $\Delta = 2$ (or $\delta = 2$) which gives the constant value of HDE in (\ref{HDE}) when (\ref{SBar}) is taken into account. In such a case, the function $\mathcal{F}$ becomes constant and the expression for dimensionless energy density (\ref{OmHDE}) just becomes 
\be
 \Omega_{HDE,\Delta=2} (a) = \frac{\rho}{\rho_{crit}} = \frac{c^2 k^2 L_0^2}{H^2(a)}  .
 \label{OmHDE2}
\ee
On the other hand, if one takes the limit $\Delta \to 2$, then one gets  from (\ref{HDEBar1}) 
\be
\Omega_{HDE}' = \Omega_{HDE} (\Omega_{HDE}-1)  \frac{\Phi'(a)}{\Phi(a)} .
\ee 

\subsection{R\'enyi nonextensive horizon entropy}

R\'enyi entropy is defined as \cite{Renyi1959}
\be
S_R(\lambda) = \frac{k_B}{1-q} \left[ \ln \sum_{i=1}^{n} p_i^q \right] = \frac{k_B}{\lambda} \ln{\left[1 + \frac{\lambda}{k_B} S_q \right]} ,
\label{S_R}
\ee
where $S_q$ is the Tsallis $q$-entropy \cite{Tsallis:1987eu}
\be
S_q =  \frac{k_B}{1-q} \left[ \sum_{i=1}^{n} p_i^q - 1 \right]
\label{S_q}
\ee
and $\lambda = 1 - q$. We will replace Tsallis $q$-entropy in (\ref{S_R}) by the dimensionless Bekenstein-Hawking entropy (\ref{SL}) in order to make it dimensionless as follows
\begin{equation}
    S_R (L,\lambda)= \frac{1}{\lambda} \ln{ \left[1+\lambda \left(\frac{L}{L_0}\right)^2 \right]} .
    \label{S_RL}
\end{equation}
For practical applications we will expand (\ref{S_RL}) into series in $\lambda$ which can be done under the assumption that the entropy deviates small from the Boltzmann-Gibbs entropy. This can be obtained from (\ref{S_R}) in the limit $\lambda \to 0$ as follows (we have added the 3rd order term for the sake of comparison with (later considered) Sharma-Mittal case only, in this subsection we perform the calculations up to the second order)
\be
S_R(L,\lambda) = S_{BH}(L) \left[ 1 - \frac{\lambda}{2} S_{BH}(L) + \ldots \right] =  \left(\frac{L}{L_0}\right)^2 \left[ 1-   \frac{\lambda}{2} \left(\frac{L}{L_0}\right)^2 + \frac{\lambda^2}{3} \left(\frac{L}{L_0}\right)^4 + \ldots \right]  
\label{SR_exp}
\ee

Applying (\ref{SR_exp}) into (\ref{omeq}), after some manipulations, one obtains 
\begin{eqnarray}
    \left(\frac{L}{L_0}\right)^2=\frac{1}{\mathcal{F}_R(\Omega_{HDE})+\frac{\lambda}{2}} ,
     \label{L_SR}
\end{eqnarray}
which further also using (\ref{L_SR}), gives the evolution equation (\ref{OmPrimeFin}) as 
\begin{eqnarray}
\Omega_{HDE}'=\frac{\Omega_{HDE}(\Omega_{HDE}-1)}{a}\left[a \frac{\Phi'}{\Phi}+\left(1-\frac{\sqrt{\Omega_{HDE}}}{k} \times \sqrt{1+\frac{\bar{\lambda}/2}{Q_R(\Omega_{HDE})}}\right)\times \left(\frac{\bar{\lambda}}{Q_R(\Omega_{HDE})}+2 \right)\right] .
\label{eveqrenyi}
\end{eqnarray}
where we rescale $\mathcal{F}(\Omega_{HDE}(a))$ from (\ref{omeq}) into 
\be
Q_R(\Omega_{HDE}) \equiv \frac{c^2}{L_0^2} \mathcal{F}_R(\Omega_{HDE}) = \frac{1}{\mathcal{A}^2} \frac{H_0^2}{k^2} \mathcal{F}_R(\Omega_{HDE}) ,
\ee
and introduced the new parameters
\begin{equation}
\bar{\lambda} \equiv \frac{\lambda c^2}{L_0^2} = \frac{\lambda}{\mathcal{A}^2} \frac{H_0^2}{k^2} .
\end{equation}
In fact, in our evaluations in Section \ref{nonext_v_data} we will find the bound on $\log(\bar{\lambda}) $. 

\subsection{Sharma-Mittal nonextensive horizon dark energy}

The Sharma-Mittal (SM) entropy \cite{SM} 
combines the R\'enyi entropy (\ref{S_R}) with the Tsallis $q-$entropy (\ref{S_q}), and is defined as
\begin{eqnarray}
    S_{SM}= \frac{k_B}{R} \left[\left(\sum_{i=1}^n (p_i)^{q}\right)^{\frac{R}{1-q}}-1 \right]  ,
    \label{SM}
\end{eqnarray}
where $R$ is another dimensionless parameter apart from $q$.  We will replace Tsallis $q$-entropy in (\ref{SM}) by the dimensionless Bekenstein-Hawking entropy (\ref{SL}) in order to make it dimensionless as follows 
\begin{eqnarray}
S_{SM}(L,\lambda,R)=\frac{1}{R}\left\{\left[1+ \lambda  S_{BH}(L) \right]^{\frac{R}{\lambda}} -1\right\} ,
\label{SSM}
\end{eqnarray}
where $R \rightarrow \lambda$ limit yields the Tsallis entropy, and $R\rightarrow 0$ limit yields R\'enyi entropy. Expanding (\ref{SSM}) into series with respect to $\lambda \approx 0$ yields
\begin{equation}
S_{SM}(L,q,p) = \left(\frac{L}{L_0}\right)^2 \left[ 1 + q  \left(\frac{L}{L_0}\right)^2 + p \left(\frac{L}{L_0}\right)^4 + \ldots \right] ,
\end{equation}
where 
\begin{equation}
q=\frac{R-\lambda}{2} ,\hspace{0.2cm} p=\frac{R^2}{6}-\frac{R \lambda}{2}+\frac{\lambda^2}{3} .
 \label{SSM_exp}
\end{equation}
Applying (\ref{SSM_exp}) into (\ref{omeq}), and performing some manipulations, we get 
\begin{eqnarray}
    \left(\frac{L}{L_0}\right)^2= \frac{(\mathcal{F}_{SM}(\Omega_{HDE})-q) \pm \sqrt{(\mathcal{F}_{SM}(\Omega_{HDE})-q)^2-4p}}{2p} ,
    \label{F_SM}
\end{eqnarray}
where $\mathcal{F}_{SM}(\Omega_{HDE})$ has been defined in (\ref{omeq}).

The evolution equation of dark energy (\ref{OmPrimeFin}) can be calculated using (\ref{F_SM}) as follows:
\begin{equation}
\Omega_{HDE}'=\frac{\Omega_{HDE}(\Omega_{HDE}-1)}{a}\left[a \frac{\Phi'}{\Phi}+\left(1-\frac{\sqrt{\Omega_{HDE}}}{k} \times \sqrt{\frac{1}{1+\bar{q} \chi +\bar{p}\chi^2} } \right)\times \left(4-\frac{2+4\bar{q}\chi+6\bar{p}\chi^2}{1+\bar{q}\chi+\bar{p}\chi^2} \right)\right] ,
\label{eveqsm}
\end{equation}
where we have defined 
\begin{equation}
\chi=\frac{(Q_{SM}(\Omega)-\bar{q}) \pm \sqrt{(Q_{SM}(\Omega)-\bar{q})^2-4\bar{p}}}{2\bar{p}} ,
\end{equation}
and 
\be
Q_{SM}(\Omega_{HDE}) \equiv \frac{c^2}{L_0^2} \mathcal{F}_{SM}(\Omega_{HDE}) = \frac{1}{\mathcal{A}^2} \frac{H_0^2}{k^2} \mathcal{F}_{SM}(\Omega_{HDE}) ,
\ee
where
\begin{equation}
\bar{q} \equiv \frac{q c^2}{L_0^2} = \frac{q}{\mathcal{A}^2} \frac{H_0^2}{k^2}= \frac{\bar{R}-\bar{\lambda}}{2} , \hspace{0.2cm} \bar{p} \equiv \frac{p c^4}{L_0^4} = \frac{1}{L_0^2}  \frac{p}{\mathcal{A}^2} \frac{H_0^2}{k^2}=\frac{\bar{R}^2}{6}-\frac{\bar{R}\bar{\lambda}}{2}+\frac{\bar{\lambda}^2}{3},
\end{equation}
and 
\begin{equation}
\bar{R}=\frac{R c^2}{L_0^2}, \bar{\lambda}=\frac{\lambda c^2}{L_0^2} .
\end{equation}
We should also mention that in our evaluations in Section \ref{nonext_v_data} we will find the bound on $\log(\bar{\lambda}) $ and $\log(\bar{R})$. 

\subsection{Kaniadakis entropy} 

Kaniadakis entropy \cite{Kaniadakis:2002zz,Kaniadakis:2005zk} results from taking into account Lorentz transformations of special relativity. It is a single $\mathcal{K}$-parameter ($-1<\mathcal{K}<1$) deformation of Boltzmann-Gibbs entropy in a similar way as is $q$-parameter for Tsallis entropy \cite{Entropy2024}, and it reads 
\be
S_{\mathcal{K}} = - k_B \sum_{i=1}^n p_i \ln_{\mathcal{K}} {p_i} =    S_k= \frac{k_B}{\mathcal{K}}\sinh\left[\mathcal{K}\frac{S}{k_B}\right] ,
\label{SKln}
\ee
where $S$ is the Boltzmann-Gibbs entropy and the definition of $K$-logarithm 
\be
\ln_K {x} \equiv \frac{x^K - x^{-K}}{2K}  
\label{lnK}
\ee
was applied. After putting the dimensionless Bekenstein-Hawking entropy  (\ref{SL}) as $S$, one gets the dimensionless form of Kaniadakis entropy as follows 
\begin{equation}
    S_k=\frac{1}{\mathcal{K}}\sinh\left(\mathcal{K}\frac{L^2}{L_0^2}\right) .
    \label{dim_Kan}
\end{equation}
Expanding (\ref{dim_Kan}) into series in small values of the parameter $\mathcal{K}$, one obtains
\begin{equation}
    S_k= \left(\frac{L}{L_0}\right)^2 \left[ 1 +\frac{\mathcal{K}^2}{6} \left(\frac{L}{L_0}\right)^2  + \ldots \right] .
    \label{K_exp}
\end{equation}
Applying (\ref{K_exp}) into (\ref{omeq}), after some manipulations, we obtain 
\begin{equation}
   \left(\frac{L}{L_0}\right)^2=\frac{\mathcal{F}(\Omega_{HDE})\pm \sqrt{\mathcal{F}^2(\Omega_{HDE})-4 \frac{\mathcal{K}^2}{6}}}{2\frac{\mathcal{K}^2}{6}} ,
    \label{K_evol}
\end{equation}
where $\mathcal{F}_{SM}(\Omega_{HDE})$ has been defined in (\ref{omeq}). This further gives the dark energy evolution equation (\ref{OmPrimeFin}) as
\begin{equation}
\Omega_{HDE}'=\frac{\Omega_{HDE}(\Omega_{HDE}-1)}{a}\left[a \frac{\Phi'}{\Phi}+\left(1-\frac{\sqrt{\Omega_{HDE}}}{k} \times  \sqrt{\frac{1}{1+\xi}} \right)\times \left(\frac{4}{1+\xi}-2 \right)\right] ,
\label{eveqka}
\end{equation}
where
\begin{equation}
\xi=\frac{1}{\frac{2}{3}\bar{\mathcal{K}}^2}\left(Q_K(\Omega_{HDE})\pm \sqrt{Q_K^2(\Omega_{HDE})-\frac{2}{3}\bar{\mathcal{K}}^2}\right)^2 ,
\end{equation}
and 
\be
Q_{K}(\Omega_{HDE}) \equiv \frac{c^2}{L_0^2} \mathcal{F}_{K}(\Omega_{HDE}) = \frac{1}{\mathcal{A}^2} \frac{H_0^2}{k^2} \mathcal{F}_{K}(\Omega_{HDE}) ,
\ee
with 
\begin{equation}
\bar{\mathcal{K}} \equiv \frac{\mathcal{K} c^2}{L_0^2} = \frac{\mathcal{K}}{\mathcal{A}^2} \frac{H_0^2}{k^2} .
\end{equation}
As in the previous cases of R\'enyi and Sharma-Mittal, we will constrain $\log(\bar{\mathcal{K}})$ in Section  \ref{nonext_v_data}. It is also important to mention that unlike R\'enyi and Sharma-Mittal, the series expansion (\ref{K_exp}) does not contain the 4th order term in the expansion, reaching the 6th order term instead. This means the influence of the nonextensivity parameter $\bar{\mathcal{K}}$ onto the evolution of the universe may in principle be even smaller than in these other cases. 

\subsection{Generalized all-in-one four-parameter and five-parameter entropic forms}

Although we do not explore these observationally, it is worth mentioning that all the above entropies are subject to a more general entropic formulas which can be recovered in the limits ot appropriate parameters. Among them there exist the four-parameter entropic formula \cite{Nojiri:2022dkr,Nojiri_4parameter}, which reads 
\be
S_g(\alpha_{\pm},\beta,\sigma,L) = \frac{1}{\sigma} \left[\left(1 + \frac{\alpha_+}{\beta} S(L) \right)^{\beta} - \left(1 + \frac{\alpha_-}{\beta} S(L) \right)^{-\beta} \right] ,
\label{4para}
\ee
as well as the five-parameter formula \cite{Odintsov:2022qnn} of the form 
\be
S_g(\alpha_{\pm},\beta,\sigma,\epsilon,L) = \frac{1}{\sigma} \left\{ \left[ 1 + \frac{1}{\epsilon} \tanh{\left( \frac{\epsilon \alpha_+}{\beta}  S(L) \right)} \right]^{\beta} -  \left[ 1 + \frac{1}{\epsilon} \tanh{\left( \frac{\epsilon \alpha_-}{\beta} S(L) \right)} \right]^{-\beta}  \right\} .
\label{5para}
\ee
When normalizing (\ref{4para}) and (\ref{5para}) by $k_B$,  the following limits allow to recover the entropies from subsections B-E above:
\begin{enumerate} 
\item if $\epsilon \to 0$, then one recovers Barrow/Tsallis-Cirto (\ref{SBar}) entropy; 
\item if $\epsilon \to 0$, $\alpha_- \to 0$, $\beta \to 0$, and $\alpha_+/\beta$ finite, then one recovers R\'enyi entropy (\ref{S_RL});
\item if $\epsilon \to 0$, $\alpha_- \to 0$, $\sigma = \alpha_+ = R$, and $\beta = R/\delta$, then one recovers Sharma-Mittal entropy formula (\ref{SSM}), though only when one replaces Tsallis $q$-entropy $S_q$ with Tsallis-Cirto $\delta$-entropy $S_{\delta}$; 
\item if $\epsilon \to 0$, $\beta \to \infty$, $\alpha_+ = \alpha_- = \sigma/2 = K$, then one recovers Kaniadakis entropy (\ref{dim_Kan}).
\end{enumerate} 

These entropies have important advantages for cosmology. Firstly, they are singular-free at the cosmological bounce (Hubble parameter $H=0$ vanishes in a bouncing scenarios \cite{Nojiri:2022aof}). Secondly, they allow microscopic interpretation \cite{Nojiri:2023bom,Nojiri:2023ikl}. Obviously, they are nonextensive.

\subsection{A generalized mass-to-horizon relation entropy}

New mass-to-horizon relation entropy has been proposed in Ref. \cite{Gohar2024} and further extended in Ref. \cite{basilakos2025}. It assumes that the mass-to-horizon-radius relation is not linear and reads 
\be
M = \gamma \frac{c^2}{G} L^n , 
\label{mass-to-hor}
\ee
where $n$ is a non-negative constant and $\gamma$ is a parameter withe dimension of $(meter)^{1-n}$. By using the Clausius relation $dE = c^2 dM = T dS$ together with the Hawking temperature definition $T = \hbar c/2 \pi k_B L$, one obtains the new entropic form 
\be
S = \gamma \frac{2n}{n+1} L^{n-1} S_{BH} ,
\ee
where $S_{BH}$ is the Bekenstein-Hawking entropy given by (\ref{BH}). By some simple rescaling in (\ref{mass-to-hor}) i.e. 
\be
L \to \frac{L}{L_0}, \hspace{0.5cm} \gamma = \tilde{\gamma} L_0 , 
\ee
one is able to convert it into 
\be
M = \tilde{\gamma} \frac{L_0 c^2}{G} \left( \frac{L}{L_0} \right)^n ,
\label{mass-to-hor_tilde}
\ee
where $\tilde{\gamma}$ is a dimensionless constant. Inserting (\ref{mass-to-hor_tilde}) into the Clausius relation, one obtains a dimensionless entropy 
\be
S_n(L) = \tilde{\gamma} \frac{2n}{n+1} \left( \frac{L}{L_0} \right)^{n-1} S_{BH}(L) , 
\label{SnL}
\ee
in which $S_{BH}(L)$ is given by (\ref{SL}) and defines the dimensionless Bekenstein-Hawking entropy. It then fits our general scheme of entropic forms comparison by inserting the formula (\ref{SnL}) into the dimensionless energy density of the holographic dark energy as given in (\ref{OmHDE}). The cosmological testing of these models has been performed in Refs.  \cite{Gohar2024,basilakos2025}.

\section{Constraining Nonextensive Holographic Dark Energy Models by Cosmological Data}
\label{nonext_v_data}

For our statistical analysis we consider various types of cosmological probes, which are detailed in the following subsections. 

The total $\chi^2$ properly corresponding to each data combination is minimized using our own code for Monte Carlo Markov Chain (MCMC). The convergence of the chains is checked using the diagnostic described in \citep{Dunkley:2004sv}. 

In order to establish the reliability of each theoretical scenario with respect to the standard $\Lambda$CDM case, we calculate the Bayes Factor \cite{doi:10.1080/01621459.1995.10476572}, $\mathcal{B}^{i}_{j}$, defined as the ratio between the Bayesian Evidences of the two compared models, in our case being the model $\mathcal{M}_j$ the reference case, $\Lambda$CDM, while the model $\mathcal{M}_i$ is the considered scenario, case by case.  We calculate the evidence numerically using our own code implementing the Nested Sampling algorithm developed by \cite{Mukherjee:2005wg}. Finally, the interpretation of the Bayes Factor is conducted using the empirical Jeffrey's scale \cite{Jeffreys1939-JEFTOP-5}.

\subsection{Pantheon+ SNeIa}

We use the Type Ia Supernovae (SNeIa) Pantheon+ sample \cite{Scolnic:2021amr,Peterson:2021hel,Carr:2021lcj,Brout:2022vxf}, made of $1701$ objects in the redshift range $0.001<z<2.26$. The $\chi^2_{SN}$ will be defined as
\begin{equation}\label{eq:chi_sn}
\chi^2_{SN} = \Delta \boldsymbol{\mathcal{\mu}}^{SN} \; \cdot \; \mathbf{C}^{-1}_{SN} \; \cdot \; \Delta  \boldsymbol{\mathcal{\mu}}^{SN} \;,
\end{equation}
where $\Delta\boldsymbol{\mathcal{\mu}} = \mathcal{\mu}_{\rm theo} - \mathcal{\mu}_{\rm obs}$ is the difference between the theoretical and the observed value of the distance modulus for each SNeIa and $\mathbf{C}_{SN}$ is the total (statistical plus systematic) covariance matrix. The theoretical distance modulus is:
\begin{equation}\label{mu_theo}
\mu_{theo}(z_{hel},z_{HD},\boldsymbol{p}) = 25 + 5 \log_{10} [ d_{L}(z_{hel}, z_{HD}, \boldsymbol{p}) ]\; ,
\end{equation}
where $d_L$ is the luminosity distance expressed in Mpc
\begin{equation}
d_L(z_{hel}, z_{HD},\boldsymbol{p})=(1+z_{hel})\int_{0}^{z_{HD}}\frac{c\,dz'}{H(z',\boldsymbol{p})} \,,
\end{equation}
with: $H(z,\boldsymbol{p})$ the Hubble parameter (cosmological model dependent through the vector of cosmological parameters $\boldsymbol{p}$); $z_{hel}$ the heliocentric redshift; $z_{HD}$ the cosmic microwave background redshift after correction from peculiar velocities \cite{Carr:2021lcj}).

The observed distance modulus $\mu_{obs}$ is $\mu_{obs} = m_{B} - \mathcal{M}$, with $m_{B}$ the standardized SNeIa blue apparent magnitude and $\mathcal{M}$ the fiducial absolute magnitude calibrated by using primary distance anchors such as Cepheids. Thanks to the $77$ SNeIa located in galactic hosts for which the distance moduli can be measured from primary anchors (Cepheids), using Pantheon+ allow to break the degeneracy between $H_0$ and $\mathcal{M}$, which can be constrained separately. Thus, the vector $\Delta\boldsymbol{\mathcal{\mu}}$ can be written as
\begin{equation}
\Delta\boldsymbol{\mathcal{\mu}} = \left\{
  \begin{array}{ll}
    m_{B,i} - \mathcal{M} - \mu_{Ceph,i} & \hbox{$i \in$ Cepheid hosts} \\
    m_{B,i} - \mathcal{M} - \mu_{theo,i} & \hbox{otherwise,}
  \end{array}
\right.
\end{equation}
with $\mu_{Ceph}$ being the Cepheid calibrated host-galaxy distance
provided by the Pantheon+ team.

\subsection{Cosmic Chronometers}

Early-type galaxies which undergo passive evolution and exhibit characteristic features in their spectra, can be defined and assessed as ``cosmic chronometers'' (CC) \cite{Jimenez:2001gg,Moresco:2010wh,Moresco:2018xdr,Moresco:2020fbm,Moresco:2022phi}, and can provide measurements of the Hubble parameter $H(z)$ \cite{Moresco:2012by,Moresco:2012jh,Moresco:2015cya,Moresco:2016nqq,Moresco:2017hwt,Jimenez:2019onw,Jiao:2022aep}. We use here the sample from \cite{Jiao:2022aep} covering the redshift range $0<z<1.965$. The corresponding $\chi^2_{H}$ can be written as
\begin{equation}\label{eq:chi_cc}
\chi^2_{H} = \Delta \boldsymbol{\mathcal{H}} \; \cdot \; \mathbf{C}^{-1}_{H} \; \cdot \; \Delta  \boldsymbol{\mathcal{H}} \;,
\end{equation}
where $\Delta \boldsymbol{\mathcal{H}} = H_{theo} - H_{data}$ is the difference between the theoretical and observed Hubble parameter, and $\mathbf{C}_{H}$ is the total (statistical plus systematics) covariance matrix calculated following prescriptions from \cite{Moresco:2020fbm}.

\subsection{Gamma Ray Bursts}

The ``Mayflower'' sample \cite{Liu:2014vda} is made of 79 GRBs in the redshift interval $1.44<z<8.1$ for which we recover the distance modulus, and overcomes the well-known issue of calibration of GRBs by relying on a robust cosmological model independent procedure. The $\chi_{G}^2$ is defined like in the SNeIa case, (\ref{eq:chi_sn}), but in this case we cannot disentangle between the Hubble constant and the absolute magnitude, so that we have to marginalize over them getting \cite{Conley:2011ku} 
\begin{equation}\label{eq:chi_grb}
\chi^2_{GRB}=a+\log d/(2\pi)-b^2/d\,,
\end{equation} 
where $a\equiv \left(\Delta\boldsymbol{\mathcal{\mu}}_{G}\right)^T \, \cdot \, \mathbf{C}^{-1}_{G} \, \cdot \, \Delta  \boldsymbol{\mathcal{\mu}}_{G}$, $b\equiv\left(\Delta \boldsymbol{\mathcal{\mu}}_{G}\right)^T \, \cdot \, \mathbf{C}^{-1}_{G} \, \cdot \, \boldsymbol{1}$ and $d\equiv\boldsymbol{1}\, \cdot \, \mathbf{C}^{-1}_{G} \, \cdot \, \boldsymbol{1}$.

\subsection{Cosmic Microwave Background}

The Cosmic Microwave Background (CMB) analysis is not performed using the full power spectra provided by \textit{Planck} \cite{Planck:2018vyg} but instead using the reduced compact likelihood based on the shift parameters defined in \cite{Wang:2007mza} and derived from the latest \textit{Planck} $2018$ data release in \cite{Zhai:2019nad}. In this case the $\chi^2_{CMB}$ can be defined as
\begin{equation}
\chi^2_{CMB} = \Delta \boldsymbol{\mathcal{F}}^{CMB} \; \cdot \; \mathbf{C}^{-1}_{CMB} \; \cdot \; \Delta  \boldsymbol{\mathcal{F}}^{CMB} \; ,
\end{equation}
where the vector $\mathcal{F}^{CMB}$ corresponds to the quantities:
\begin{eqnarray} \label{eq:CMB_shift}
R(\boldsymbol{p}) &\equiv& \sqrt{\Omega_m H^2_{0}} \frac{r(z_{\ast},\boldsymbol{p})}{c}, \nonumber \\
l_{a}(\boldsymbol{p}) &\equiv& \pi \frac{r(z_{\ast},\boldsymbol{p})}{r_{s}(z_{\ast},\boldsymbol{p})}\,,
\end{eqnarray}
in addition to a constraint on the baryonic content, $\Omega_b\,h^2$, and on the dark matter content, $(\Omega_m-\Omega_b)h^2$. In (\ref{eq:CMB_shift}), $r_{s}(z_{\ast},\boldsymbol{p})$ is the comoving sound horizon evaluated at the photon-decoupling redshift. The general definition of the comoving sound horizon is
\begin{equation}\label{eq:soundhor}
r_{s}(z,\boldsymbol{p}) = \int^{\infty}_{z} \frac{c_{s}(z')}{H(z',\boldsymbol{p})} \mathrm{d}z'\, ,
\end{equation}
where the sound speed is given by
\begin{equation}\label{eq:soundspeed}
c_{s}(z) = \frac{c}{\sqrt{3(1+\overline{R}_{b}\, (1+z)^{-1})}} \; ,
\end{equation}
with the baryon-to-photon density ratio parameters defined as $\overline{R}_{b}= 31500 \Omega_{b} \, h^{2} \left( T_{CMB}/ 2.7 \right)^{-4}$ and $T_{CMB} = 2.726$ K. The photon-decouping redshift is evaluated using the fitting formula from \cite{Aizpuru:2021vhd}, accurate up to $ \sim 0.0005\%$,
\begin{equation}{\label{eq:zdecoupl}}
    z_{\ast} =
    \frac{391.672 \omega_m^{-0.372296}  + 937. 442 \omega_b^{-0.97966}}{\omega_m^{-0.0192951}\omega_b^{-0.93681}} + \omega_m^{-0.731631}\, ,
\end{equation}
with $\omega_m = \Omega_{m}\,h^2$ and $\omega_b = \Omega_{b}\,h^2$. 

Finally, $r(z_{\ast}, \boldsymbol{p})$ is the comoving distance at decoupling, i.e. using the definition of the comoving distance:
\begin{equation}\label{eq:comovdist}
d_{M}(z,\boldsymbol{p})=\int_{0}^{z} \frac{c\, dz'}{H(z',\boldsymbol{p})} \; ,
\end{equation}
we set $r(z_{\ast},\boldsymbol{p}) = d_M(z_{\ast},\boldsymbol{p})$.

\subsection{Baryon Acoustic Oscillations} 

For BAO we consider multiple data sets from different surveys. In general, the $\chi^2$ is defined as
\begin{equation}
\chi^2_{BAO} = \Delta \boldsymbol{\mathcal{F}}^{BAO} \, \cdot \ \mathbf{C}^{-1}_{BAO} \, \cdot \, \Delta  \boldsymbol{\mathcal{F}}^{BAO} \ ,
\end{equation}
with the observables $\mathcal{F}^{BAO}$ which change from survey to survey.

The WiggleZ Dark Energy Survey \cite{Blake:2012pj} gives, at redshifts $z=\{0.44, 0.6, 0.73\}$, the acoustic parameter
\begin{equation}\label{eq:AWiggle}
A(z,\boldsymbol{p}) = 100  \sqrt{\Omega_{m} \, h^2} \frac{d_{V}(z,\boldsymbol{p})}{c \, z} \, ,
\end{equation}
where $h=H_0/100$, and the Alcock-Paczynski distortion parameter
\begin{equation}\label{eq:FWiggle}
F(z,\boldsymbol{p}) = (1+z)  \frac{d_{A}(z,\boldsymbol{p})\, H(z,\boldsymbol{p})}{c} \, ,
\end{equation}
where $d_{A}$ is the angular diameter distance defined as
\begin{equation} \label{eq:ang_dist}
d_{A}(z,\boldsymbol{p})=\frac{1}{1+z}\int_{0}^{z} \frac{c\, dz'}{H(z',\boldsymbol{p})} \;,
\end{equation}
and
\begin{equation}
d_{V}(z,\boldsymbol{\theta})=\left[ (1+z)^2 d^{2}_{A}(z,\boldsymbol{\theta}) \frac{c z}{H(z,\boldsymbol{\theta})}\right]^{1/3}
\end{equation}
is the geometric mean of the radial and tangential BAO modes.

The final release of the Sloan Digital Sky Survey (SDSS) Extended Baryon Oscillation Spectroscopic Survey (eBOSS) observations \cite{Tamone:2020qrl,deMattia:2020fkb,BOSS:2016wmc,Gil-Marin:2020bct,Bautista:2020ahg,Nadathur:2020vld,duMasdesBourboux:2020pck,Hou:2020rse,Neveux:2020voa}, with the exception of the SDSS-IV DR14 quasars analysis from \cite{Zhao:2018gvb}, provides:
\begin{equation}
\frac{d_{M}(z,\boldsymbol{p})}{r_{s}(z_{d},\boldsymbol{p})}, \qquad \frac{c}{H(z,\boldsymbol{p}) r_{s}(z_{d},\boldsymbol{p})} \,,
\end{equation}
where the comoving distance $d_M$ is given by (\ref{eq:comovdist}) and the sound horizon is evaluated at the dragging redshift $z_{d}$. The dragging redshift is estimated using the analytical approximation, accurate up to $\sim 0:001\%$, provided in  \cite{Aizpuru:2021vhd} which reads
\begin{equation}\label{eq:zdrag}
    z_d =
    \frac{1 + 428.169 \omega_b^{0.256459} \omega_m^{0.616388} + 925.56 \omega_m^{0.751615}}{\omega_m^{0.714129}}\, ,
\end{equation}
with $\omega_m = \Omega_m\, h^2$ and $\omega_b = \Omega_b\, h^2$. 

Data from \cite{Zhao:2018gvb} are expressed in terms of 
\begin{equation}
d_{A}(z,\boldsymbol{p}) \frac{r^{fid}_{s}(z_{d},)}{r_{s}(z_{d},\boldsymbol{p})}, \qquad H(z,\boldsymbol{p}) \frac{r_{s}(z_{d},\boldsymbol{p})}{r^{fid}_{s}(z_{d},\boldsymbol{p})} \,,
\end{equation}
where $r^{fid}_{s}(z_{d})$ is the sound horizon at dragging redshift calculated for the given fiducial cosmological model considered in \cite{Zhao:2018gvb}, which is equal to $147.78$ Mpc.

Finally, we have added the most recent BAO data collected by the Dark Energy Spectroscopic Instrument (DESI), as reported in \cite{DESI:2024mwx}. BAO data are given at seven different redshift values, ranging from $0.295$ to $2.330$, and are related to various types of tracers (Bright Galaxies, Luminous Red Galaxies, Emission Line Galaxies, Quasars and Lyman-$\alpha$ Forest). Depending on the tracer, DESI provides measurement of the previously defined quantities
\begin{equation}
\frac{d_{M}(z,\boldsymbol{p})}{r_{s}(z_{d},\boldsymbol{p})}, \qquad \frac{c}{H(z,\boldsymbol{p}) r_{s}(z_{d},\boldsymbol{p})},  \qquad \frac{d_{V}(z,\boldsymbol{p})}{r_{s}(z_{d},\boldsymbol{p})} \,,
\end{equation}
as illustrated in Table~1 of \cite{DESI:2024mwx}.

{\renewcommand{\tabcolsep}{1.5mm}
{\renewcommand{\arraystretch}{2.}
\begin{table*}
\begin{minipage}{\textwidth}
\caption{Comparative MCMC analysis of $\Lambda$CDM, Tsallis-Cirto, Barrow, Kaniadakis, R\'enyi, standard HDE (Bekenstein) (cf. formula (\ref{stHDE})), and Sharma-Mittal holographic dark energy models. We report $1\sigma$ confidence intervals for each parameter and the Bayes Factors. Italic font is used for cases in which the posterior is noisy or not properly limited. In those cases we provide the interval defined by $\Delta \chi^2 = \chi^2 - \chi_{min}^2 = 1$ and the value at the minimum of the $\chi^2$ in brackets. The value of Barrow parameter $\Delta$ is underlined since it falls outside the range of its applicability for Tsallis-Cirto parameter $\delta$ (cf. (\ref{Deldel})).}
\label{tab:results_integrated}
\centering
\resizebox*{\textwidth}{!}{
\begin{tabular}{c|ccccccccc}
\hline
   & $\Omega_m$ & $\Omega_b$ & $h$ & $\mathcal{M}$ & $k$ & $\Delta$ / $\delta$ & $\log \bar{\lambda} $ or $\log \bar{\cal{K}}$& $\log{\bar{R}}$ & $\log \mathcal{B}^{i}_{j}$ \\
\hline
\hline
$\Lambda$CDM & $0.297^{+0.004}_{-0.004}$ & $0.0476^{+0.0005}_{-0.0005}$ & $0.686^{+0.004}_{-0.003}$ & $-19.41^{+0.01}_{-0.01}$ & - & - & - & - & $0$ \\
\hline
Tsallis-Cirto & $0.298^{+0.005}_{-0.004}$ & $0.0480^{+0.0006}_{-0.0005}$ & $0.683^{+0.004}_{-0.004}$ & $-19.41^{+0.01}_{-0.01}$ & $138^{+48}_{-57}$ & $>\underline{1.86}/1.93$ & - & - & $-0.56^{+0.03}_{-0.03}$ \\
\hline
Barrow & $0.292^{+0.005}_{-0.005}$ & $0.0479^{+0.0009}_{-0.0008}$ & $0.685^{+0.005}_{-0.005}$ & $-19.42^{+0.01}_{-0.01}$ & $5.33^{+0.89}_{-1.22}$ & $>0.86/1.43$ & - & - & $-5.91^{+0.04}_{-0.04}$ \\
\hline
Kaniadakis & $0.287^{+0.005}_{-0.005}$ & $0.0481^{+0.0008}_{-0.0008}$ & $0.684^{+0.005}_{-0.005}$ & $-19.43^{+0.01}_{-0.01}$ &  $0.73^{+0.02}_{-0.03}$ & - & $\mathit{[0.26,2.75] (1.25)}$ & - & $-14.64^{+0.03}_{-0.03}$ \\
\hline
R\'enyi & $0.287^{+0.005}_{-0.005}$ & $0.0481^{+0.0008}_{-0.0007}$ & $0.684^{+0.005}_{-0.006}$ & $-19.42^{+0.01}_{-0.01}$ & $0.73^{+0.03}_{-0.03}$ & - & $\mathit{<0.25 (-432)}$ &  - & $-14.67^{+0.03}_{-0.03}$ \\
\hline
HDE (Bekenstein) & $0.287^{+0.005}_{-0.005}$ & $0.0481^{+0.0008}_{-0.0008}$ & $0.685^{+0.005}_{-0.005}$ & $-19.42^{+0.01}_{-0.01}$ & $0.73^{+0.03}_{-0.03}$ & - & - & - & $-14.69^{+0.03}_{-0.03}$ \\
\hline
Sharma-Mittal & $0.287^{+0.005}_{-0.005}$ & $0.0481^{+0.0009}_{-0.0008}$ & $0.684^{+0.006}_{-0.005}$ & $-19.43^{+0.01}_{-0.01}$ & $0.73^{+0.03}_{-0.03}$ & - & $\mathit{[-0.31,1.08] (0.63)}$  & $\mathit{[-2.95,1.89] (1.87)}$ &  $-14.69^{+0.03}_{-0.03}$ \\
\hline
\end{tabular}}
\end{minipage}
\end{table*}}}

 \section{Discussion and Conclusions}
\label{Discussion}

The bounds on the cosmological parameters for all the considered scenarios are given in Table~\ref{tab:results_integrated}.  We constrain the following parameters: dimensionless $k$ which applies for all the models starting from standard HDE (Bekenstein entropy), $\delta$ and $\Delta$ for Tsallis-Cirto and Barrow entropies: $\bar{\lambda}$ for R\'enyi entropy, $\bar{R}$ and $\bar{\lambda}$ for Sharma-Mittal entropy, and  $\bar{\mathcal{K}}$ for Kaniadakis entropy. 

The first main consideration relates to the Bayes factor $\mathcal{B}^{i}_{j}$. Indeed it is clear that, given Jeffreys' scale $\ln \mathcal{B}^{i}_{j}$, all the alternative models considered here are statistically disfavoured with respect to $\Lambda$CDM, although not all with at the same level. For example, we can see how the Kaniadakis, R\'enyi, Holographic DE (Bekenstein-Hawking), and Sharma-Mittal models are very strongly disfavoured, with a logarithm of the Bayes factor $\sim -14.7$. This is a quite strong results, for which we can undoubtedly state that they can be rejected as possible alternative to the standard cosmological constant.

If we focus on these models and the corresponding values of the cosmological parameters, we can see how all parameters are basically consistent with those found for the $\Lambda$CDM case, except for $\Omega_m$, which differs at the $2\sigma$ level. But even this difference cannot really be considered as the main reason for such a strong incompatibility with the data. Actually, it may be seen as the explicit expression of the intrinsic impossibility of these models to replicate the cosmological background dynamics, not even by twisting the values of the cosmological parameters. It is also important that the underlying HDE model, which is basically Bekenstein-Hawking entropy model, is excluded and it is are definitely strongly nonextensive due to its area-entropy (not volume-entropy) framework. All three other models seem to be some small modifications of HDE models due to the series expansions in their extra parameter $\lambda$. For what concerns the characteristic parameters of these models ($k$, $\lambda$, $\mathcal{K}$, $R$), we can further notice how $k$ is quite well constrained, and that there is basically no difference from one scenario to another. 
The other parameters, instead, are basically unconstrained with a tendency to prefer very small values (R\'enyi scenario), or they are limited to a specific range of values, but the posterior we get from our MCMC is quite noisy and no clear median (and errors) can be derived from them. In such cases, we provide a rough estimation of the interval in which $\Delta \chi^2 = \chi^2 - \chi_{min}^2 = 1$ and the value at the $\chi_{min}^2$ within brackets. We have verified that this behaviour is due to a very negligible role of the correction(s) produced by this parameter on the dynamical equations. We also add that this does not seem to be related to the choice of working with Taylor series expansions. Indeed, we have also verified that if we try to perform a numerical inversion of (\ref{omeq}), in order to get a relation between the entropy models, the sound horizon and the parameters, we arrive to the same conclusions, i.e. the role of $\lambda$ is minimal, if not negligible at all.

Some more interesting conclusions can be driven for the Tsallis-Cirto and the Barrow models. We remind here that the two, although with different physical motivations, might be considered as belonging to one single framework, once the relation (\ref{Deldel}) between their two parameters (respectively, $\delta$ for Tsallis-Cirto and $\Delta$ for Barrow) is made explicit.

For the Barrow case, we can see how the logarithm of the Bayes Factor still penalizes this scenario with respect to the reference case, although in much lesser way that the previously discussed models. We note that the addition of DESI BAO data with respect to the data sets used in our previous works, worsen the fit of the data, with the logarithm of the Bayes factor moving from $\sim-3$ to $\sim -6$.

What is actually more interesting in the context of the literature about this scenario, is that we confirm the results we found in our previous works \cite{Dabrowski:2020atl,PhysRevD.108.103533}: the Barrow fractal index $\Delta$ is bound from below ($\Delta > 0.86$ and peaking towards $\Delta \to 1$), thus implying that the cosmological horizon should be of the fractal nature. This value equivalent to Tsallis-Cirto $\delta > 1.43$ is very close to the value of $\delta = 3/2$ ($\Delta = 1$) recently claimed in Ref. \cite{TSALLIS2025139238} as best-fitting value giving the extensivity of entropy (though still violating additivity) and so preserving the Legendre structure of thermodynamics. This has a very interesting consequence in view of the nonextensive entropies cosmological applications - namely, the fit to the data is possible for minimal nonextensivity in the system. Otherwise, the resolution of the dark energy problem due to the Barrow holography is possible if the thermodynamics of the universe is more alike the classical Boltzmann-Gibbs in view of the extensivity, but not the additivity (see the discussion in Ref. \cite{Entropy2024}). However, still there are subtleties since 
for the entropic form to be extensive, the form should be homogeneous function of order one of the thermodynamic quantities \cite{Entropy2024}. In $\Delta\rightarrow 1$ case, this is true for length (the entropy scales with volume), but not for the internal energy  \cite{Tsallis2025}. 
In view of the results of this paper, the ``standard'' limit of a non-fractal horizon, i.e. $\Delta = 0$, is excluded by our set of data and, in fact, our bound on Barrow index contradicts the evaluations coming from baryon asymmetry for Tsallis-Cirto entropy of Ref. \cite{Luciano2022a} being $\mid \delta - 1 \mid \approx 10^{-3}$ and for Barrow entropy \cite{LucianoSar} being $0.005 \leq \Delta \leq 0.008$ as well as the big-bang nucleosynthesis evaluation for Tsallis-Cirto being $1 - \delta < 10^{-5}$ \cite{Lambiase2021}, power-law inflation \cite{Luciano2023} and combined cosmological data evaluation of Ref. \cite{Sheykhi2021} of $\delta = 0.9999$ (marked with parameter $\beta$ in this Ref.). It also contradicts the bounds from the perturbations in Ref. \cite{PhysRevD.99.103524} where both Tsallis models THDE1 and THDE2 give the value of $\delta$ close to unity i.e. $\delta = 0.939$ and $\delta = 0.941$. 

Finally, we have the Tsallis-Cirto model, which can be seen as (almost) fully equivalent to the $\Lambda$CDM case. We point out how the same $\Lambda$CDM, corresponding to $\delta = \Delta = 2$, is actually the limiting value of the fit, as we obtain $\delta >1.93$ (or $\Delta > 1.86$) peaking towards $\Delta \to 2$. We stress that we cannot exactly recover the $\delta = 2$ value by using Eq. (\ref{HDETsallisC}), because in this limit they are not valid anymore. On the other hand, the value of $\Delta =2$ is beyond the range of this Barrow parameter and also cannot be applied. However, still it is clear how our data points us towards this value which is more suitable for the standard $\Lambda$CDM model. Though potentially, the dark energy evolution equations (\ref{HDEBarfin}) and   (\ref{HDETsallisC}) look identical while looking at (\ref{Deldel}), in our MCMC calculations, we have applied the prior for Barrow parameter $0 \leq \Delta \leq 1$ (the limit of maximum fractality) which is not the case for Tsallis-Cirto parameter $\delta$ which resulted in formally two different bounds (if Barrow parameter was considered extended) rather than one bound, in the table \ref{tab:results_integrated}. 

It is also worth saying that our results favour the lower value of the Hubble parameter and in that sense the Hubble tension is not alleviated. This is unlike in the recent investigation of Ref. \cite{Basilakos2024}. The point is that in our approach we apply holographic principle rather than gravity-thermodynamics conjecture. In the latter, the cosmological constant is already present and the horizon entropy contribution allows only for the ''correction'' of the standard $\Lambda$CDM model which leads to the small values of the nonadditivity parameter (in the Tsallis-Cirto case $\delta \approx 1$; equivalent to $\Delta \approx 0$ in the Barrow case). In our approach we do not allow cosmological constant trying to fully substitute it by the HDE from the horizon entropy which leads to high values of nonextensivity parameters. The full discussion of the problem was given in our previous paper \cite{PhysRevD.108.103533}. 

Last remark is that any model which does not have a power law correction (i.e. Renyi, Sharma-Mittal, Kaniadakis) yields the holographic dark energy model, where $k$ is well constrained and the non extensivity parameters are really small or not constrained at all. For model which contains power law corrections (Barrow and Tsallis-Cirto) the data prefers $\Lambda$CDM and picks values of nonextensivity parameter that converts the model to $\Lambda$CDM. This may also explain why the parameter $k$ is much larger ($k \sim 138$ instead of the order of one) for the Tsallis-Cirto model compared to Barrow and other models - it is just trying to place itself closer to $\Lambda$CDM by gaining from the $k$ holographic purposes rather than using nonextensivity given by $\delta$ value.



\section*{Acknowledgments}
The work of I.C. and M.P.D. was supported by the Polish National Science Centre grant No. DEC-2020/39/O/ST2/02323. All the Authors acknowledge the discussions with Constantino Tsallis and Hussain Gohar. 

\bibliographystyle{apsrev4-1} 
\bibliography{HDE_19}

\end{document}